\documentstyle[epsfig,12pt]{article}

\newcommand{\bce}{\begin{center}} 
\newcommand{\ece}{\end{center}}
\newcommand{\beq}{\begin{equation}}
\newcommand{\eeq}{\end{equation}}
\newcommand{\bea}{\vspace{0.25cm}\begin{eqnarray}}
\newcommand{\eea}{\end{eqnarray}}

\newcommand{\ba}{\begin{array}}
\newcommand{\ea}{\end{array}}

\newcommand{\rhs}{{\sl rhs~}}

\newcommand{\doublespace}{
    \renewcommand{\baselinestretch}{1.6}\large\normalsize}

\def\lsim{\mathrel{\rlap{\lower4pt\hbox{\hskip1pt$\sim$}}
    \raise1pt\hbox{$<$}}}         
\def\gsim{\mathrel{\rlap{\lower4pt\hbox{\hskip1pt$\sim$}}
    \raise1pt\hbox{$>$}}}         

\def\beq{\begin{equation}}
\def\endeq{\end{equation}}
\def\arr{\begin{eqnarray}}
\def\endarr{\end{eqnarray}}
\makeindex

  \advance\oddsidemargin  by -1in
  \advance\evensidemargin by -1in
  \advance\topmargin      by -0.5in
\begin{document}

\vspace{2.0cm}

\vspace{1.0cm}

\begin{center}
{\Large \bf 
High energy neutrino in a nuclear environment:\\
mirror asymmetry of the shadowing effect}\footnote{Talk presented at 
XXXIII International Conference on High Energy Physics
26.07.06-02.08.06 Moscow}

\vspace{1.0cm}
V.R.~Zoller

\vspace{1.0cm}

{\it
ITEP, Moscow 117218, Russia\\}
\vspace{1.0cm}
{ \bf Abstract }\\
\end{center}
The parity non-conservation effect in diffractive  charged current DIS is 
quantified in terms of color dipole sizes of left-handed and right-handed 
electroweak bosons.
We identify the origin  and estimate the strength of the left-right 
asymmetry effect   and present
comparison with experimental data on the parity-odd structure function 
$\Delta xF_3 =xF_3^{\nu N}-xF_3^{\bar\nu N}$.
We study the shadowing effect in absorption 
 of left-handed  and right-handed 
$W$-bosons by  atomic nuclei. 
The target nucleus is found to be quite  transparent for
the charmed-strange  Fock component  of the light-cone
$W^+$  in the helicity state $\lambda=+1$
and  rather opaque for the $c\bar s$ dipole  with  $\lambda=-1$.

\doublespace

\vskip 0.5cm \vfill $\begin{array}{ll}
\mbox{{\it email address:}} & \mbox{zoller@itep.ru} \\
\end{array}$

\pagebreak



In this paper we discuss the effects of mirror asymmetry in small-$x$
charged current (CC) deep inelastic scattering (DIS). 
Since the physical picture 
of a process may change dramatically in different reference frames, 
one can gain deeper insight into the dynamics of a process by choosing a
 particular frame. In a frame where the parton picture of a nucleon is manifest
the mirror asymmetry appears as a difference of relevant parton and 
anti-parton
densities, in a dual frame the mirror asymmetry shows up as a difference 
of color dipole sizes of left-handed and right-handed $W$-bosons 
($W_{L,R}$). 
Indeed, in the brick wall frame  in the chiral 
limit the $W_L$ interacts only with quarks while the $W_R$ interacts only 
with anti-quarks. Hence, the representation for the standard parity-odd 
structure function $F_3$ in terms of parton 
densities  \cite{Ioffe}
\bea
F_3^{\nu p}=2\{(d-{\bar u})+(s-{\bar c})\} \\
F_3^{\nu n}=2\{(u-{\bar d})+(s-{\bar c})\}\\
F_3^{{\bar\nu} p}=2\{(u-{\bar d})+(c-{\bar s})\}\\
F_3^{{\bar\nu} n}=2\{(d-{\bar u})+(c-{\bar s})\} \,,
\label{eq:F3PN}
\eea
The isospin symmetry of the nucleon sea 
implies that at small Bjorken $x$ 
the  structure function $F_3$ is dominated by 
the charm-strange   weak current 
and for an isoscalar nucleon target at $x\to 0$
\bea
F_3^{\nu N}=2(s-{\bar c}) \\
F_3^{{\bar\nu} N}=2(c-{\bar s})\,.
\label{eq:F3ISO}
\eea
The non-partonic sea contribution to $F_3$ 
 coming from
the  u-d current is  proportional to the light quark mass splitting  
$\sim \Delta m_{ud}/m_d$ and vanishes in the limit $m_u=m_d$.

In the dipole/laboratory  frame the space-time picture of the 
phenomenon is quite different \cite{FZ1,FZ2}.
In the color dipole approach \cite{NZ91, M} 
(for the review see \cite{HEBECKER})  the
  small-$x$ DIS is treated in terms of the
interaction of the quark-antiquark color dipole of size ${\bf r}$,
that 
the virtual $W$ transforms into,  
 with the target nucleon.  This interaction is described by the beam- 
 and flavor-independent color dipole cross section
$\sigma(x,r)$. At small $x$ the dipole size ${\bf r}$ 
 is a conserved quantum number and the parity non-conservation effect can be 
quantified in terms of color dipole sizes of left-handed and right-handed 
W-bosons. Hence, the representation of $F_3$ in terms of
absorption cross sections for $W_L$- and $W_R$-bosons ($\sigma_{L,R}$)
\beq
F_3\sim \sigma_L-\sigma_R\,.
\label{eq:F3SIGMA}
\eeq 
These cross sections
 are calculated as a quantum mechanical expectation values of
$\sigma(x,r)$.
Once the light-cone wave function (LCWF) of a color dipole
state is specified the evaluation of $\sigma_{L,R}$
 becomes a routine quantum mechanical procedure.

Below  we report our extension of  the color dipole analysis
 onto the charged current   DIS  with particular emphasis 
on the left-right 
 asymmetry  of diffractive
interactions of electroweak bosons of different helicity.
We make use of the  LCWF derived in \cite{FZ1,FZ2} and evaluate the structure
 functions $xF_3$  in the vacuum exchange dominated region
of $x\lsim 0.01$. We present comparison of our results with experimental data. 
In experiments with neutrino beams the nuclear targets are in use and 
 the nucleon structure functions are  extracted 
from  nuclear data. In particular,  from the differential  cross section 
of $\nu Fe$-scattering. 
The latter is considered as the  incoherent sum 
of $\nu N$ cross sections. 
However, at small $x$ the effect of non-additivity of nuclear  cross sections
plays important role. The interference of multiple scattering amplitudes 
leads to $\sigma^{\nu A}\not= A\sigma^{\nu N}$ and   
to the so called  nuclear shadowing effect.
The shadowing effect is found to be different for left-handed and 
right-handed $W$-bosons and  mirror asymmetry is shown to be enhanced by
 the large  thickness of a 
nucleus.

At small $x$
the contribution of diffractive excitation of open 
charm/strangeness to the absorption cross section for scalar, $(\lambda=0)$,
left-handed, $(\lambda=-1)$,  and right-handed, $(\lambda=+1)$,  
$W$-boson of virtuality $Q^2$,
is given by the color dipole
factorization formula \cite{ZKL,BBGG}
\beq
\sigma_{\lambda}(x,Q^{2})
=\int dz d^{2}{\bf{r}} \sum_{\lambda_1,\lambda_2}
|\Psi_{\lambda}^{\lambda_1,\lambda_2}(z,{\bf{r}})|^{2} 
\sigma(x,r)\,.
\label{eq:FACTORN}
\eeq
In Eq.~(\ref{eq:FACTORN}) $\Psi_{\lambda}^{\lambda_1,\lambda_2}(z,{\bf{r}})$
 is the LCWF of
the $|c\bar s\rangle$ state with the $c$ quark 
carrying fraction $z$ of the $W^+$ light-cone momentum and 
$\bar s$ with momentum fraction $1-z$ \cite{FZ1,FZ2}. 
The $c$- and $\bar s$-quark
helicities are  $\lambda_1=\pm 1/2$ and  $\lambda_2=\pm 1/2$, respectively.

Only diagonal elements of  the density matrix \cite{FZ1,FZ2}
\beq
\rho_{\lambda\lambda^{\prime}}
=\sum_{\lambda_1,\lambda_2}\Psi_{\lambda}^{\lambda_1,\lambda_2}
\left(\Psi_{\lambda^{\prime}}^{\lambda_1,\lambda_2}\right)^*
\label{eq:RHO}
\eeq
for $\lambda=\lambda^{\prime}=L,R$ 
enter Eq.~(\ref{eq:FACTORN}): 
\bea
\rho_{RR}(z,{\bf r})=\left|\Psi_{R}^{+1/2, +1/2}\right|^2
+\left|\Psi_{R}^{-1/2, +1/2}\right|^2\nonumber\\
={{8\alpha_W N_c}\over (2\pi)^2}(1-z)^2\left[m^2 K^2_0(\varepsilon r)
+\varepsilon^2 K^2_1(\varepsilon r)\right]
\label{eq:RHOR}
\eea
and
\bea
\rho_{LL}(z,{\bf r})=\left|\Psi_{L}^{-1/2,-1/2}\right|^2
+\left|\Psi_{L}^{-1/2,+1/2}\right|^2\nonumber\\
={{8\alpha_W N_c}\over (2\pi)^2}z^2\left[\mu^2 K^2_0(\varepsilon r)
+\varepsilon^2 K^2_1(\varepsilon r)\right],
\label{eq:RHOL}
\eea
where
\beq
\varepsilon^2=z(1-z)Q^2+(1-z)m^2+z\mu^2\,,
\label{eq:VAREP}
\eeq
$K_{\nu}(x)$ is the modified Bessel function, $\alpha_W={g^2/4\pi}$ and 
\beq
{G_F\over \sqrt{2}}={g^2\over m^2_{W}}.
\label{eq:GF}
\eeq  
The $c$-quark 
and $\bar s$-quark masses are $m$ and $\mu$, respectively.

The left-handedness of weak currents leads to 
the striking momentum partition asymmetry of both  
$\rho_{LL}$ and $\rho_{RR}$ \cite{FZ1,FZ2}.
The left-handed quark in  the decay of  left-handed $W^+_L$ carries away 
most  of the $W^+_L$ light-cone momentum.The same does the right-handed
antiquark in the light-cone decay of $W_R^+$.
This is the   phenomenon of the same nature as  
 spin-spin correlations in the neutron $\beta$-decay.  
The observable which is 
  strongly affected by this left-right asymmetry  is the parity-odd
structure function of the neutrino-nucleon DIS named $F_3$. Its definition
in terms of $\sigma_{R}$ and $\sigma_{L}$ of Eq.~(\ref{eq:FACTORN}) 
is as follows:
\beq
2xF_3(x,Q^2)={Q^2\over 4\pi^2\alpha_{W}}
\left[\sigma_{L}(x,Q^{2})-\sigma_{R}(x,Q^{2})\right],
\label{eq:F3}
\eeq

To estimate consequences of the left-right asymmetry for
$F_3$ at high $Q^2$, such that 
\beq
{m^2\over Q^2}\ll 1,\,\,{\mu^2\over Q^2}\ll 1~,  
\label{eq:DLLA}
\eeq
one should take into account that
 the dipole cross-section $\sigma(x,r)$ in 
Eq.~(\ref{eq:FACTORN}) is related to the un-integrated
gluon structure function
${{\cal F}(x,\kappa^2)}={\partial G(x,\kappa^2)/\partial\log{\kappa^2}},
$ \cite{GLUE}:
\bea
\sigma(x,r)={\pi^2 \over N_c}r^2\alpha_S(r^2)
\int{d\kappa^2\kappa^2\over (\kappa^2+\mu_G^2)^2}
{4[1-J_0(\kappa r)]\over \kappa^2r^2}{{\cal F}(x_g,\kappa^2)}~.
\label{eq:SIGMAN}
\eea
In the Double Leading Logarithm Approximation (DLLA), 
i.e. for small dipoles,
\bea
\sigma(x,r)\approx {\pi^2\over N_c} r^2\alpha_S(r^2)
G(x_g,A/r^2),
\label{eq:SMALL}
\eea
where $\mu_G=1/R_c$ is the inverse correlation radius of perturbative gluons
and $A\simeq 10$ comes from properties of the Bessel function $J_0(y)$.
Because of scaling violation $G(x,Q^2)$  rises with $Q^2$, but the product
$\alpha_S(r^2)G(x,A/r^2)$ is approximately flat in $r^2$.
At large $Q^2$ the leading
contribution to $\sigma_{\lambda}(x,Q^{2})$ comes from
the P-wave term, $\varepsilon^2 K_1(\varepsilon r)^2$, in
Eqs.~(\ref{eq:RHOR}) and (\ref{eq:RHOL}).
 The asymptotic behavior of the 
Bessel function, $K_1(x)\simeq \exp(-x)/\sqrt{2\pi/x}$ makes the 
$\bf{r} $-integration rapidly convergent at $\varepsilon r > 1$.
 Integration over $\bf{r}$ in Eq.~(\ref{eq:FACTORN}) yields
\beq
\sigma_L\propto \int_0^1 dz {z^2\over \varepsilon^2}\alpha_SG\sim 
{\alpha_SG\over Q^2}\log{Q^2\over \mu^2}
\label{eq:SIGL}
\eeq
and similarly 
\beq
\sigma_R\propto \int_0^1 dz {(1-z)^2\over \varepsilon^2}\alpha_SG
\sim  {\alpha_SG\over Q^2}\log{Q^2\over m^2}.
\label{eq:SIGR}
\eeq
It should be
emphasized that we focus on  the vacuum exchange contribution 
to $xF_3$ corresponding
to the excitation of the $c\bar s$ state in the process 
 the $W^+$-gluon fusion, 
\beq
W^+g\to c\bar s. 
\label{eq:WG}
\eeq
Therefore, the structure function $xF_3$ 
differs from zero only  due to the strong left-right asymmetry of the 
light-cone $|c\bar s\rangle$ Fock state.
One should bear in mind, however, that in experiment
 the smallest available values of $x$ 
are in fact only moderately small and there is 
quite significant  valence contribution to $xF_3$.
 The valence term, $xV$ is the same
for both  $\nu N$ and 
$\bar\nu N$ structure functions of an  iso-scalar nucleon. 
The sea-quark term in the $xF^{\nu N}_3$ which we are interested in is
 denoted by $xS(x,Q^2)$ and  has opposite
sign for $xF^{\bar\nu N}_3$. Indeed,  the substitution 
$m\leftrightarrow\mu$ in Eqs.~(\ref{eq:RHOR}) and (\ref{eq:RHOL}) entails
$\sigma_L\leftrightarrow\sigma_R$.
 Therefore,  
\beq
xF^{\nu N}_3=xV+xS,
\label{eq:SVNU}
\eeq
and
\beq
xF^{\bar\nu N}_3=xV-xS.
\label{eq:SVANU}
\eeq

\begin{figure}[ht]
\psfig{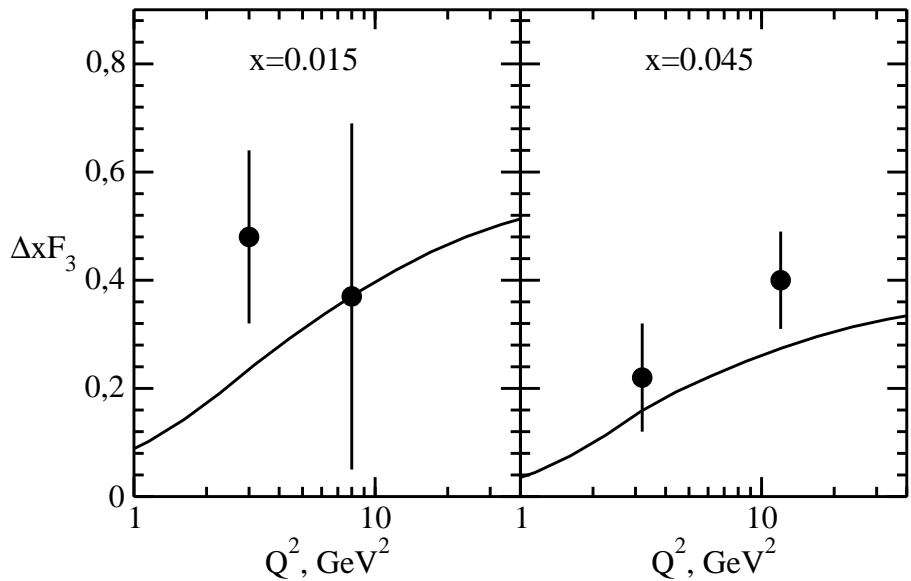}
\caption{$\Delta x F_3$ data  as a function of $Q^2$ \cite{CCFR3}.
 Shown by  solid lines are the results of  color dipole description.} 
\label{fig:fig1}
\end{figure}

One can combine the $\nu N$ and 
$\bar\nu N$ structure functions to isolate the Pomeron exchange term,
\beq
\Delta x F_3=xF_3^{\nu N}-xF_3^{\bar \nu N}=2xS.
\label{eq:fig2}
\eeq
The extraction of $\Delta x F_3$
from CCFR $\nu_{\mu}Fe$ and $\bar\nu_{\mu}Fe$ differential cross section
in a model-independent way has been reported in \cite{CCFR3}. Figure 
\ref{fig:fig1} shows  the extracted values of $\Delta x F_3$
as a function of  $Q^2$  for two 
smallest values of $x$.
The solid curves are calculated  making use of Eqs.~(\ref{eq:FACTORN}) and 
(\ref{eq:SIGMAN}) with the  differential
gluon density function
${{\cal F}(x_g,\kappa^2)}$ determined in \cite{NIKIV}.
The  $\Delta x F_3$ 
slowly increases with growing $Q^2$ because 
of the logarithmic scaling violation in $\sigma_{L,R}$.
Note that in   the region of moderately small $x$, $x>0.01$,
 the mass threshold effect
suppresses $\Delta x F_3 $ at $Q^2\lsim (m+\mu)^2$

The CCFR/NuTeV structure functions 
$xF^{\nu N}_3$ and $xF^{\bar\nu N}_3$ are  extracted from the 
$\nu Fe$  and $\bar\nu Fe$ data \cite{CCFR,CCFR3}. 
For the nuclear  absorption cross section the color dipole
factorization gives
\bea
\sigma^A_{\lambda}(x,Q^{2})
=\langle \Psi_{\lambda}| \sigma^A(x,r)|\Psi_{\lambda}\rangle\nonumber\\
=\int dz d^{2}{\bf{r}} \sum_{\lambda_1,\lambda_2}
|\Psi_{\lambda}^{\lambda_1,\lambda_2}(z,{\bf{r}})|^{2} 
\sigma^A(x,r)\,,
\label{eq:FACTORA}
\eea
where \cite{GLAUBER}
\beq
\sigma^A(x,r)=
2\int d^{2}{\bf{b}}
\left\{1-\exp\left[-{1\over 2}\sigma(x,r)T(b)\right]\right\}. 
\label{eq:SIGA}
\eeq
Here $T(b)$ is the the optical  thickness of a nucleus,  
\beq 
T(b)=\int_{-\infty}^{+\infty} dz n(\sqrt{z^2+b^2}),
\label{eq:TB}
\eeq 
${\bf b}$ is the impact parameter and $n(r)$ is the nuclear matter density 
normalized as follows: 
\beq
\int d^3r n(r)=A.
\label{eq:NA}
\eeq 
It is assumed that $A\gg 1$. 
One can expand the exponential in Eq.~(\ref{eq:SIGA}) to  separate  
the impulse approximation term 
and the  shadowing correction, $\delta\sigma^A_{\lambda}$, 
in  (\ref{eq:FACTORA}),
\beq
\sigma^A_{\lambda}=
A\sigma_{\lambda}
-\delta\sigma^A_{\lambda}.
\label{eq:NUCLEAR}
\eeq

To the lowest order  in $\sigma T$ the shadowing term  reads 
\beq
\delta\sigma^A_{\lambda}\simeq
{\pi\over 4} \langle \sigma_{\lambda}^2\rangle
{\cal S}^2_A(k_L)\int db^2 T^2(b),
\label{eq:SHADOW}
\eeq
where
\bea
\langle \sigma_{\lambda}^2\rangle=
\langle\Psi_{\lambda}|\sigma(x,r)^2|\Psi_{\lambda} \rangle\nonumber\\
={\cal S}^2_A(k_L)\int dz d^{2}{\bf{r}} \sum_{\lambda_1,\lambda_2}
|\Psi_{\lambda}^{\lambda_1,\lambda_2}(z,{\bf{r}})|^{2} 
\sigma^2(x,r).
\label{eq:SIG2AV}
\eea
The longitudinal nuclear form factor ${\cal S}_A(k_L)$ in
Eq.~(\ref{eq:SHADOW}) takes care about the coherency constraint,
\beq
l\gg R_A.
\label{eq:COHCON}
\eeq
The approximation (\ref{eq:SHADOW}) represents the driving term of shadowing, 
the double-scattering term. It is reduced by the higher-order rescatterings
by about $30\%$ for iron  and $50\%$ for lead nuclei. This 
 accuracy is  quite sufficient   for  order-of-magnitude  estimates. 
The numerical calculations presented below are done for the full Glauber series
(\ref{eq:SIGA}),  
\beq
\delta\sigma^A_{\lambda}=\pi{\cal S}^2_A(k_L)\sum_{n=2}^{\infty}
{(-1)^n\langle \sigma_{\lambda}^n\rangle\over n! 2^{n-1}}
\int db^2 T^n(b)\,,
\label{eq:GSERIES}
\eeq 
where the effect of finite coherence length is modeled by the factor
${\cal S}^2_A(k_L)$ in  $\rhs$. A consistent description of the latter
effect in electro-production was obtained in Ref.~\cite{BGZ98} based on  the 
light-cone path integral technique of  Ref.~\cite{BGZ96}.

 To estimate  the strength of the nuclear shadowing effect in
$xF_3$ at high $Q^2$ such that 
\beq
{m^2\over Q^2}\ll 1,\,\,{\mu^2\over Q^2}\ll 1  
\label{eq:DLLAA}
\eeq
one can use  the dipole cross section $\sigma(x,r)$ of Eq.(\ref{eq:SIGMAN})
Let us estimate first the
contribution to $\langle\sigma^2_{\lambda}\rangle$ coming from
the P-wave term, $\varepsilon^2 K_1(\varepsilon r)^2$, in Eqs.~(\ref{eq:RHOR})
and (\ref{eq:RHOL}).  Integration over $\bf{r}$ in Eq.~(\ref{eq:SIG2AV}) yields
\beq
\langle\sigma^2_{L}\rangle\propto \int_0^1 dz {z^2\over \varepsilon^4}\propto 
{1\over Q^2 \mu^2}
\label{eq:SIGLA}
\eeq
and similarly 
\beq
\langle\sigma^2_{R}\rangle\propto \int_0^1 dz {(1-z)^2\over \varepsilon^4}
\propto  {1\over Q^2m^2}.
\label{eq:SIGRA}
\eeq
Obviously, the integral (\ref{eq:SIGLA}) is dominated by $z\gsim 1-\mu^2/Q^2$
i.e., by $\varepsilon^2\sim \mu^2$ and, consequently, by 
$r^2\sim 1/\varepsilon^2\sim  1/\mu^2$. A comparable  contribution
to (\ref{eq:SIGLA}) comes from the S-wave term
$\propto \mu^2 K_0(\varepsilon r)^2$ in $\rho_{LL}$. In Eq.~(\ref{eq:SIGRA})
the integral
is dominated by  $z\lsim m^2/Q^2$, corresponding to 
$\varepsilon^2\sim m^2$. Therefore, 
$r^2\sim 1/\varepsilon^2\sim  1/m^2$.
Thus, we conclude that the 
typical dipole sizes which dominate $\sigma_{\lambda}$ and
$\langle \sigma_{\lambda}^2\rangle$ are very different.
In Ref.~\cite{FZ1} basing on the color dipole approach  we found 
the scaling cross sections 
$\sigma_L$ and $\sigma_R$ $\propto 1/Q^2$
times the Leading-Log scaling violation factors
$\propto\log Q^2/\mu^2$ and $\propto\log Q^2/m^2$, respectively 
(see Eqs.(\ref{eq:SIGL},\ref{eq:SIGR})). 
The scaling violations  were found to be  (logarithmically) dominated by 
\beq 
r^2\sim 1/Q^2.
\label{eq:1Q2}
\eeq
On the contrary, the contribution of small-size  dipoles,
$\sim 1/Q^2$, to $\langle \sigma_{\lambda}^2\rangle$,  defined in
Eq.~(\ref{eq:SIG2AV}), proved to be  negligible. 
At $\lambda=-1$ $\langle \sigma_{\lambda}^2\rangle$
 is dominated by large hadronic size $c\bar s$-dipoles, $ r\sim 1/\mu$. 
Consequently,  
\beq 
\delta\sigma^A_L\propto  1/\mu^2.
\label{eq:S2L}
\eeq
At $\lambda=+1$ a typical $c\bar s$-dipole  is rather  small, $r\sim 1/m$,
and $\delta\sigma^A_R$ is small as well:
\beq 
\delta\sigma^A_R\propto 1/m^2.
\label{eq:S2R}
\eeq

\begin{figure}[ht]
\psfig{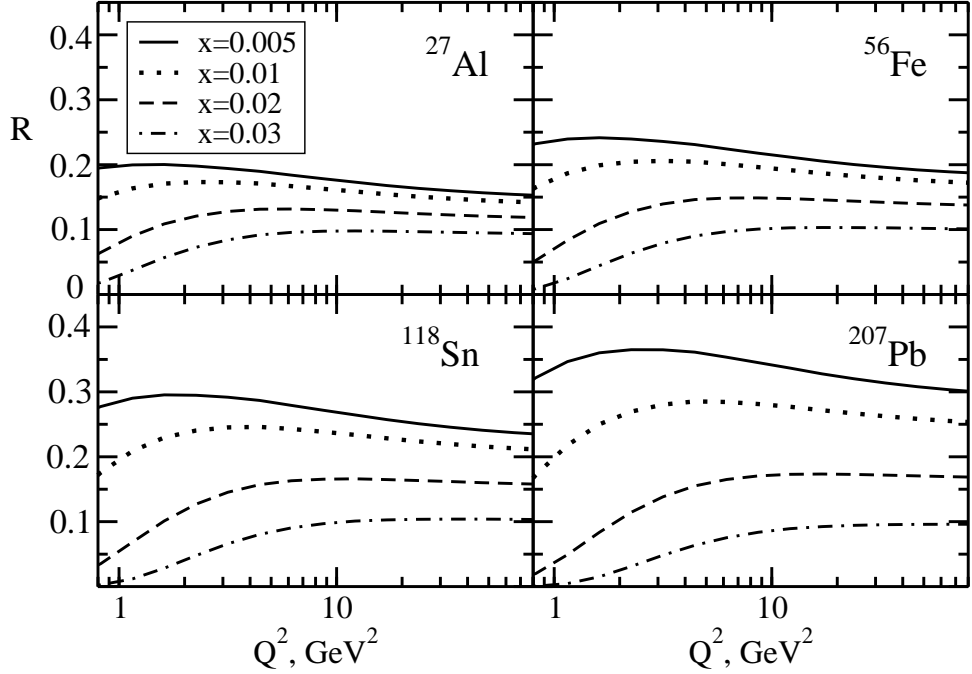}
\vspace{-0.5cm}
\caption{The shadowing ratio $R$ as a function of $Q^2$ for 
several values of $x$
 calculated from the nuclear charge 
densities of Ref.\cite{DEVRIES} for some sample nuclei.}
\label{fig:RNUC}
\end{figure} 

 Thus, there is a sort of filtering phenomenon,
the target nucleus absorbs the $c\bar s$ Fock component of $W^+$
 with $\lambda=-1$, but 
is nearly transparent for $c\bar s$ states with opposite helicity,
  $\lambda=+1$.  

From Eq.(\ref{eq:NUCLEAR}) it follows that  
the shadowing correction to nucleonic $\Delta xF_3$ extracted 
from nuclear data is
\beq
\delta(\Delta xF_3)={Q^2\over 4\pi^2\alpha_W}{1\over A}
\left(\delta \sigma^A_{L}-\delta \sigma^A_{R} \right).
\label{eq:DELF3}
\eeq 
To give an idea of the magnitude of the shadowing
 effect we evaluate the ratio of the nuclear shadowing correction,
$\delta(\Delta x F_3)$,
 to the 
nuclear structure function of the impulse approximation, $A\cdot\Delta x F_3$,
\beq
R={\delta(\Delta x F_3)\over A\Delta x F_3}=
{\delta\sigma^A_L-\delta\sigma^A_R\over
A\sigma_L-A\sigma_R}.
\label{eq:R}
\eeq
 Hence, the nuclear shadowing correction  
$R\cdot\Delta  xF_3$
which should be added to  $\Delta xF_3$ extracted from
the $\nu Fe$ data to get the  ``genuine'' $\Delta xF_3$. 
This correction positive-valued  and does
increase $\Delta xF_3$ of the impulse approximation.
We calculate $R$ as a function of $Q^2$ for several values 
of Bjorken $x$ in the 
kinematical range of CCFR/NuTeV experiment. 
Our results obtained for  realistic nuclear densities of 
Ref.\cite{DEVRIES} are presented 
 in Figure \ref{fig:RNUC}.  Shown is  
the ratio $R(Q^2)$ for different nuclear 
targets including $^{56}Fe$. 
At small $x$ and  high $Q^2$  the shadowing correction
scales, $\delta\sigma_{L,R}\propto 1/Q^2$.
 The absorption cross section $\sigma_{L,R}$ 
scales as well. The ratio $\delta\sigma_{L,R}/\sigma_{L,R}$ 
slowly decreases with growing $Q^2$ because 
of the logarithmic scaling violation in $\sigma_{L,R}$.
Toward the region of $x>0.01$,
 both the nuclear form factor and the mass threshold effect
suppress $R$ at $Q^2\lsim (m+\mu)^2$ (see Fig.\ref{fig:RNUC}).

Summarizing, we 
developed the light-cone color dipole
description  of the left-right asymmetry effect 
in charged current DIS at small Bjorken $x$. 
We evaluated the contribution of the diffractive excitation of
charm-strange Fock states of the light-cone W-boson to  the structure function 
$\Delta x F_3=xF_3^{\nu}-xF_3^{\bar \nu}$ and 
compared our results with experimental data. 
 Theory is in reasonable agreement with data.
We  presented  the color dipole analysis of nuclear 
effects in charge current DIS. The emphasis was put on
the pronounced effect of  left-right  asymmetry of shadowing in
neutrino-nucleus DIS at small values of Bjorken $x$. 
We predicted strikingly
different scaling behavior of nuclear shadowing for the left-handed and
right-handed $W^+$. Large, about $20-25\%$, shadowing in the 
$Fe$ structure functions
is predicted, which is important for a precise determination of
the nucleon structure functions $xF_3$ and $\Delta xF_3$. 

\vspace{0.2cm} \noindent \underline{\bf Acknowledgments:}

The work was supported in part by the RFBR grant 06-02-16905-a.

\end{document}